# Dirac Nodal Lines and Tilted Semi-Dirac Cones Coexisting in a Striped Boron Sheet

Honghong Zhang,[†] Yuee Xie,[†] Zhongwei Zhang,[†] Chengyong Zhong,[†]

Yafei Li,[‡] Zhongfang Chen,[#] Yuanping Chen[†,*]

[†] School of Physics and Optoelectronics, Xiangtan University, Xiangtan, Hunan, 411105, China

[‡] College of Chemistry and Materials Science, Jiangsu Key Laboratory of Biofunctional Materials, Nanjing Normal University, Nanjing, Jingsu, 210023, China.

[#] Department of Chemistry, Institute for Functional Nanomaterials, University of Puerto Rico, Rio Piedras Campus, San Juan, PR 00931, USA

To whom correspondence should be addressed. Email:    chenyp@xtu.edu.cn

**ABSTRACT:**

The enchanting Dirac fermions in graphene stimulated us to seek for other two-dimensional (2D) Dirac materials, and boron monolayers may be a good candidate. So far, a number of monolayer boron sheets have been theoretically predicted, and three have been experimentally prepared. However, none of them possesses Dirac electrons. Herein, by means of density functional theory (DFT) computations, we identified a new boron monolayer, namely hr-sB, with two types of Dirac fermions coexisting in the sheet: one type is related to Dirac nodal lines traversing Brillouin zone (BZ) with velocities approaching $10^6$ m/s, the other is related to tilted semi-Dirac cones with strong anisotropy. This newly predicted boron monolayer consists of hexagon and rhombus stripes. With an exceptional stability comparable to the experimentally achieved boron sheets, it is rather optimistic to grow hr-sB on some suitable substrates such as the Ag (111) surface. The unique electronic properties induced by special bond characteristics also imply that this boron monolayer may be a good superconductor.

**Graphical Abstract**

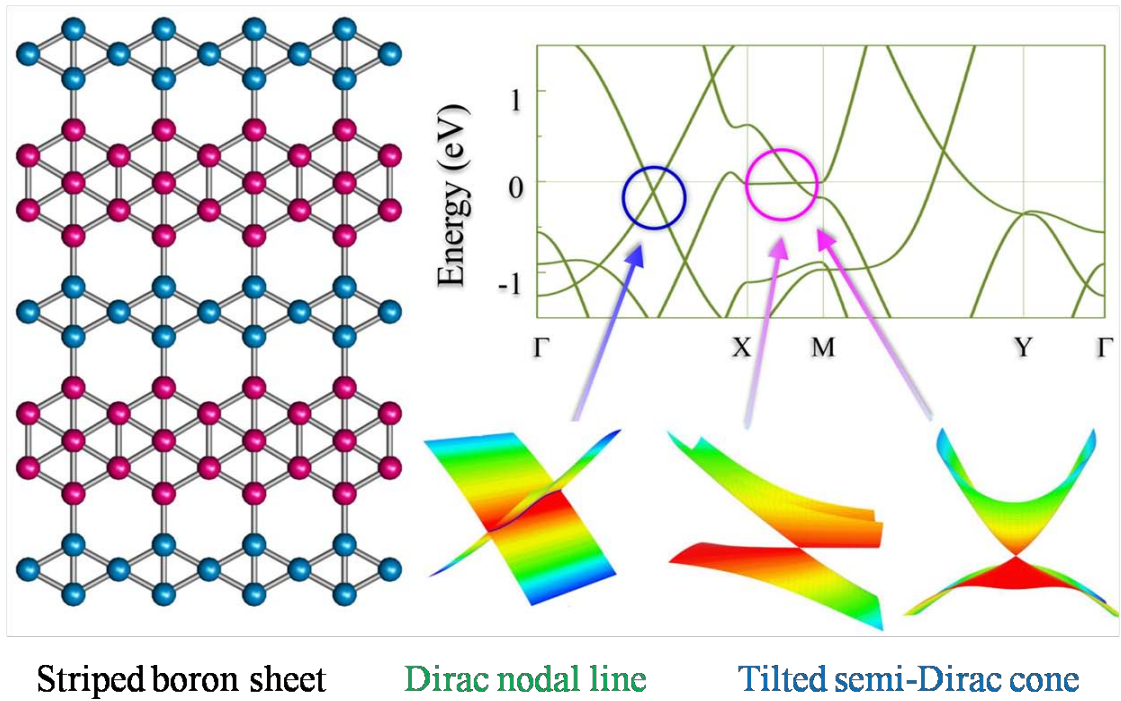

A striped boron sheet (hr-sB) with unique electronic properties: coexisting of Dirac nodal lines and tilted semi-Dirac cones.

**Introduction**

Graphene, a monolayer carbon sheet, has been one of the hottest materials in the last decade. Possessing massless Dirac fermions is an import reason for graphene's excellent electronic properties.[1-6] Besides graphene, Dirac electrons have also been found in other monolayer carbon sheets, such as graphyne.[7-9] Finding new physical phenomena behind Dirac electrons and extending related applications are strongly driving our efforts to seek for other two-dimensional (2D) materials with massless Dirac fermions. Especially, the Dirac materials with Dirac nodal lines or tilted Dirac cones (type-II Dirac cones) have been the new focus of this field.[10-14]

As the left neighbor of carbon in the periodic table, boron is possibly the second element after carbon that can have multiple 2D allotropes.[15-19] So far, a number of *monolayer* boron sheets have been theoretically predicted, such as *α*-, *β*-, *γ*-, and *δ*-type boron sheets.[15, 20] The atoms in the sheets appreciate to form triangular or hexagonal rings. Therefore, all of these boron sheets can be constructed by inserting atoms to honeycomb lattice or removing atoms from triangular lattice.[15, 20, 21] A parameter, namely hexagonal vacancy density η, has been introduced to describe the ratio of hexagon holes to the number of atomic sites in the original triangular lattice within one unit cell.[15, 22, 23] The electronic properties of the boron sheets are correlated to the value of η.[15, 16, 24] Inspiringly, three boron monolayer structures have been experimentally achieved. Mannix *et al.* successfully synthesized the triangular boron sheet $\delta_6$ (η=0) on silver surfaces under ultrahigh-vacuum conditions,[25] while Feng *et al.* demonstrated that $\beta_{12}$- (η=1/6) and $\chi_3$-sheets (η=1/5) can be grown epitaxially on a

Ag (111) substrate.[26] With the development of fabrication techniques, it is expected that more monolayer boron sheets can be realized in the very near future.

Though boron and carbon are neighbors in the periodic table, their electronic properties and bonding characters are thoroughly different. In the 2D carbon sheets, carbon atoms usually form sp$^2$ hybridization,[27-31] though sp hybridization is also possible.[7-9, 32] In most cases, p$_z$ orbital dominates the electron states on the Fermi level, and thus the Dirac electrons around the Fermi level are induced by the interactions of p$_z$ orbitals.[7, 9, 27, 29, 32-35] In contrast, the electron deficient boron tends to form multiple center bonds in the 2D boron allotropes, and the p$_z$ orbital is not the only dominating orbital any more.[15, 18, 23, 36] Thus, it is rather hard to form Dirac cones in 2D boron monolayers due to complicated bands around the Fermi level. Consequently, Dirac electrons have never been found in the monolayer boron sheets up to date, only a few multi-layer allotropes or buckled structures functioned by hydrogen (or oxygen) atoms possess Dirac cones.[37-42] A question arises naturally, is there an experimentally feasible monolayer boron sheet with Dirac fermions?

In this paper, by means of density functional theory (DFT) computations, we successfully identified a new boron sheet consisting of hexagon as well as rhombus stripes, which has an exceptional stability and unique Dirac fermions. This new boron allotrope has a high stability comparable to the experimentally available $\delta_6$-, $\beta_{12}$- and $\chi_3$- sheets. Interestingly, Dirac nodal lines and tilted semi-Dirac cones coexist around the Fermi level. The two Dirac nodal lines traverse the whole Brillouin zone (BZ), and the Dirac points in the nodal lines are crossed by two linear bands, corresponding to two one-dimensional (1D) channels in the hexagon and rhombus stripes, respectively, while the two tilted semi-Dirac cones are at the sides of the BZ, and are featured tilted axis and anisotropic band crossings. Our chemical bonding analyses showed that this boron sheet has very special bonding patterns, and a tight-binding model well describes different orbital interactions in the structure. Additionally, we proposed a possible method to synthesize this new boron monolayer.

**Model and Computational Methods**

Figure 1 presents the optimized structure of our designed boron sheet, which features stripes of hexagons and rhombi along *x* axis, and thus named as hexagon- and rhombus-stripe boron (hr-sB). According to the stripes, the boron atoms can be divided into two kinds: the $B_h$ atoms in the hexagon stripes and the $B_r$ atoms in the rhombus stripes. Note that all the $B_r$ atoms are four-fold coordinated, while the $B_h$ atoms are four-, five- or six-fold coordinated, respectively. This boron allotrope is purely planar without any buckling. Its primitive cell is a rectangle including 5 $B_h$ and 3 $B_r$ atoms, respectively (as shown by the dashed lines in Figure 1). The hexagonal vacancy density ($\eta=1/5$) is the same as that of $\chi_3$-sheet.

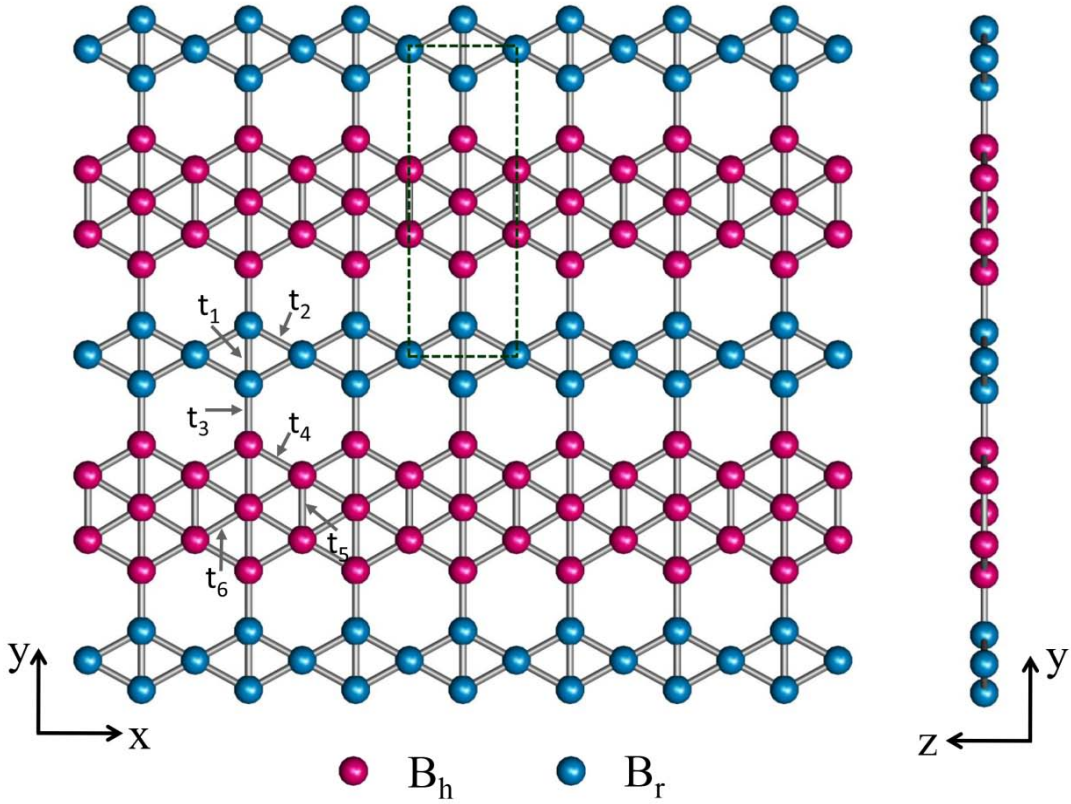

**Figure 1.** The top and side views of the optimized structure of hr-sB. The boron atoms in hexagon and rhombus are denoted as the $B_h$ and $B_r$, and colored as wine and blue, respectively. $t_1 \sim t_6$ represent the hopping energies between different atoms. The unit cell is shown as dashed line.

All the DFT computations were performed using Vienna ab *initio* simulation program package (VASP).[43] The Perdew-Burke-Ernzerhof (PBE) functional was employed for the exchange-correlation term according to generalized gradient approximation (GGA).[44, 45] The projector augmented wave (PAW) method[46] was used to represent the ion-electron interaction, and the kinetic energy cutoff of 400 eV was adopted. The atomic positions were fully optimized by the conjugate gradient method, the energy and force convergence criteria were set to be $10^{-5}$ eV and $10^{-3}$ eV/Å, respectively. A vacuum region of 10 Å was added to avoid interaction between adjacent images. The BZ was sampled using 13×5×1 Monkhorst-Pack $k$-points meshes.[47] The phonon calculations were carried out using the Phonopy package[48] with the forces calculated by the VASP code. To evaluate the thermal stability, we carried out ab initio molecular dynamics (AIMD) simulations with canonical ensemble, for which a 3 × 7 supercell containing 168 atoms was used and the AIMD simulations were performed with a Nose-Hoover thermostat at 700 and 800 K, respectively. The Fermi velocity was calculated with the expression $v_F=E/\hbar k$, where the $E/k$ is the slope of linear valence band (VB) or conduction band (CB) and the ℏ is the reduced Plank's constant.

**Results and discussions**

Table 1 summarizes the structural parameters and the energetic data of hr-sB, and the corresponding properties of the experimentally available $\delta_6$-, $\beta_{12}$- and $\chi_3$- boron sheets for comparison. hr-sB is purely planar, and its lattice constants are $a = 2.91$ Å and $b = 8.38$ Å along the $x$ and $y$ axes, respectively. Its bond lengths vary from 1.61 to 1.76 Å, closing to those of $\beta_{12}$- and $\chi_3$- sheets. The same as $\beta_{12}$- sheet, hr-sB also has the space group of *Pmmm*.

**Table1.** The lattice constants, bond lengths, space group, planar or buckling, and cohesive energy ($E_c$) for the experimentally synthesized $\beta_{12}$, $\chi_3$ and $\delta_6$, and the newly predicted hr-sB in this work. All results are calculated based on PBE functional.

| Structures | $a$ (Å) | $b$ (Å) | Bond lengths (Å) | Space group | Planar or buckling | $E_c$ (eV/atom) |
|---|---|---|---|---|---|---|
| $\delta_6$ | 2.87 | 1.61 | 1.61-1.89 | Pmmn | buckling | -6.183 |
| hr-sB* | 2.91 | 8.38 | 1.61-1.76 | Pmmm | planar | -6.190 |
| $\beta_{12}$ | 2.92 | 5.07 | 1.65-1.75 | Pmmm | planar | -6.231 |
| $\chi_3$ | 4.45 | 4.45 | 1.62-1.72 | Cmmm | planar | -6.245 |

\* this work

The cohesive energy of hr-sB (6.19 eV/atom) is comparable with the experimentally available $\delta_6$ (6.18 eV/atom), $\beta_{12}$ (6.23 eV/atom) and $\chi_3$ (6.24 eV/atom) boron sheets, thus it has rather high thermodynamic stability. No soft modes were found in the computed phonon dispersion (Figure 2a), implying its dynamic stability. We also examined its thermal stability by performing AIMD simulations in canonical ensemble. After heating up to the targeted temperature 700 K for 20 *ps*, we did not observe any structural decomposition (Figure 2b). However, the structure reconstruction occurs at 800 K during the 20 *ps* simulation. These computational results show that the newly predicted hr-sB nanosheet has rather sound thermodynamic stability and outstanding dynamic and thermal stabilities, thus are highly feasible for experimental realization.

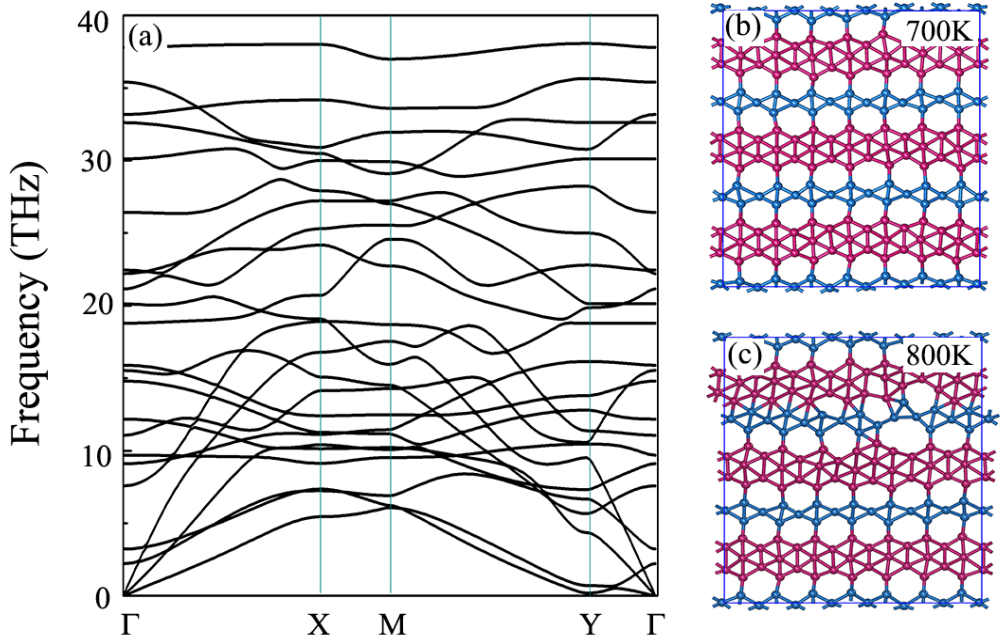

**Figure 2.** (a) The phonon dispersion of the hr-sB. Snapshots for the equilibrium structures of hr-sB monolayer at the temperatures of (b) 700K, (c) 800K, after 20 *ps* AIMD simulations.

Interestingly, the band structure of hr-sB exhibits anisotropic electronic properties (Figure 3). Along the *k* path Γ − X, two linear bands (labeled as D1 and D2) cross together and form a Dirac point around the Fermi level. The same case occurs along M − Y. A closer examination reveals that the linear band-crossing in fact occurs along two lines traversing the BZ along $k_y$: one resides around $k_x \approx 0.25\,\pi/a$, and the other resides around $-0.25\,\pi/a$ because of time reversal or inversion symmetry, as shown in Figure 3c. The three-dimensional (3D) band spectrum (Figure 3d) illustrates that the nodal line is formed by crossing of two planes. The Dirac fermions transport along $k_x$, and the velocities can approach $(4.6 \sim 10.2) \times 10^5$ *m/s*.

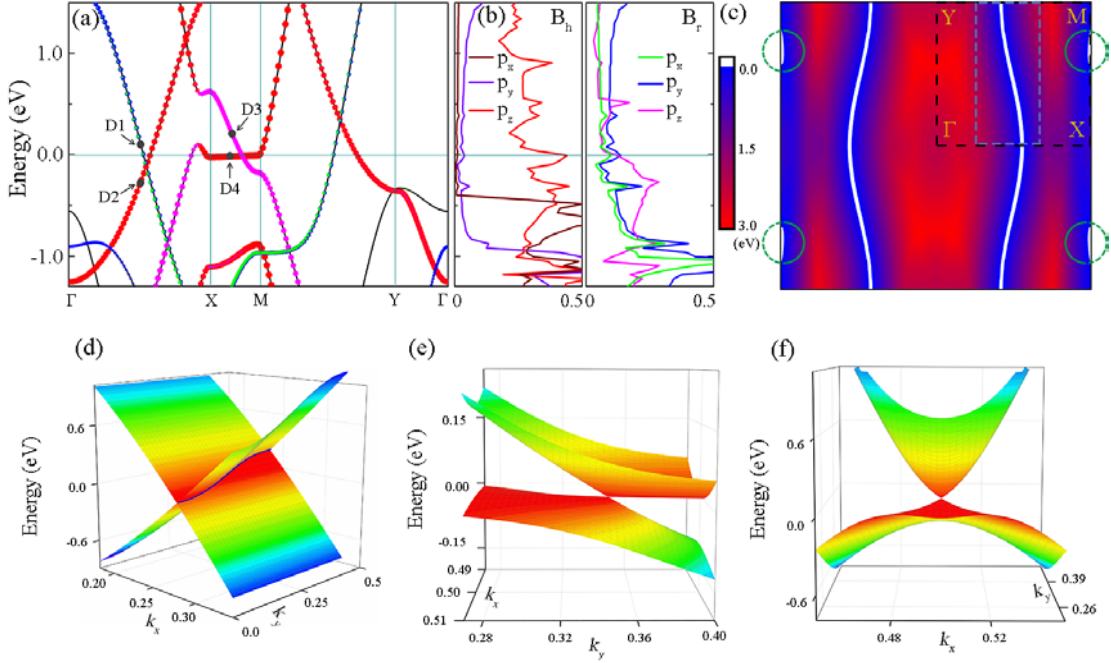

**Figure 3.** (a) The projected band structures of hr-sB. (b) PDOS for the $B_h$ and $B_r$ atoms, respectively. (c) A schematic figure of Dirac nodal line in 2D BZ. The color bar indicates the energy difference between conduction band and valence band at each $k$ points. (d) 3D band structures in the vicinity (labeled as cyan dash line in (c)) of the nodal line. (e)-(f) 3D plots of the band structures around the crossing point of bands D3 and D4 in (a).

In contrast, along the $k$ path X − M, a linear band D3 and a flat band D4 cross at the Fermi level (Figure 3a). The electron velocity corresponding to the band D3 is 3.7 × $10^5$ $m/s$, while that corresponding to the band D4 is almost zero. To reveal the nature of the crossing point, we plotted the 3D band spectra in Figures 3e and 3f. Figure 3e shows that it is a distorted Dirac cone having a tilted center axis, somewhat similar to the type-II Dirac cone.[10-12] However, Figure 3f illustrates that, along the $k_x$ direction, the crossing point is a contact of two quadratic bands rather than linear bands. Therefore, this is a tilted semi-Dirac cone, which represents a kind of new fermions never been observed in other structures. According to the time reversal or inversion symmetry, there are two tilted semi-Dirac cones distributing at the sides of the first BZ (see the circles in Figure 3c).

To explore the origin of the exotic band crossings, we calculated the projected density of state (PDOS) (Figure 3b). Clearly the electron states around the Fermi level are attributed by different orbitals in $B_h$ and $B_r$ atoms. For $B_h$ atoms, almost only $p_z$ orbital has contributions to the states, while for $B_r$ atoms, all the $p$ orbitals ($p_x$, $p_y$ and $p_z$) have significant contributions. Such orbital contributions are also reflected in the computed band structure (Figure 3a). It is known that, in graphene, the two crossing linear bands both come from $p_z$ orbitals. In contrast, in hr-sB sheet, the two crossing bands come from different orbitals in different atoms. The band D2 is from $p_z$ orbital of $B_h$ atom, while the other band D1 is from $p_{x/y}$ orbital of $B_r$ atom.

To compare the states in the two bands, we plotted the charge densities of states in bands D1 and D2 around the Fermi level in Figure 4. For the D1 state, the $p_{x/y}$ orbitals form σ bands along the rhombus stripes, while for the D2 state, the $p_z$ orbitals form π bands along the hexagon stripes. Because the orbital interactions between $p_{x/y}$ orbitals of $B_r$ atoms and $p_z$ orbitals of $B_h$ atoms are very weak, the two "stripe states" are almost isolated, in other words, there exists 1D channels for Dirac electrons in the rhombic and hexagonal stripes, respectively. The two bands, corresponding to the two 1D channels, cross together and form a Dirac point, which extends to a Dirac nodal line along the periodic direction perpendicular to the stripes.

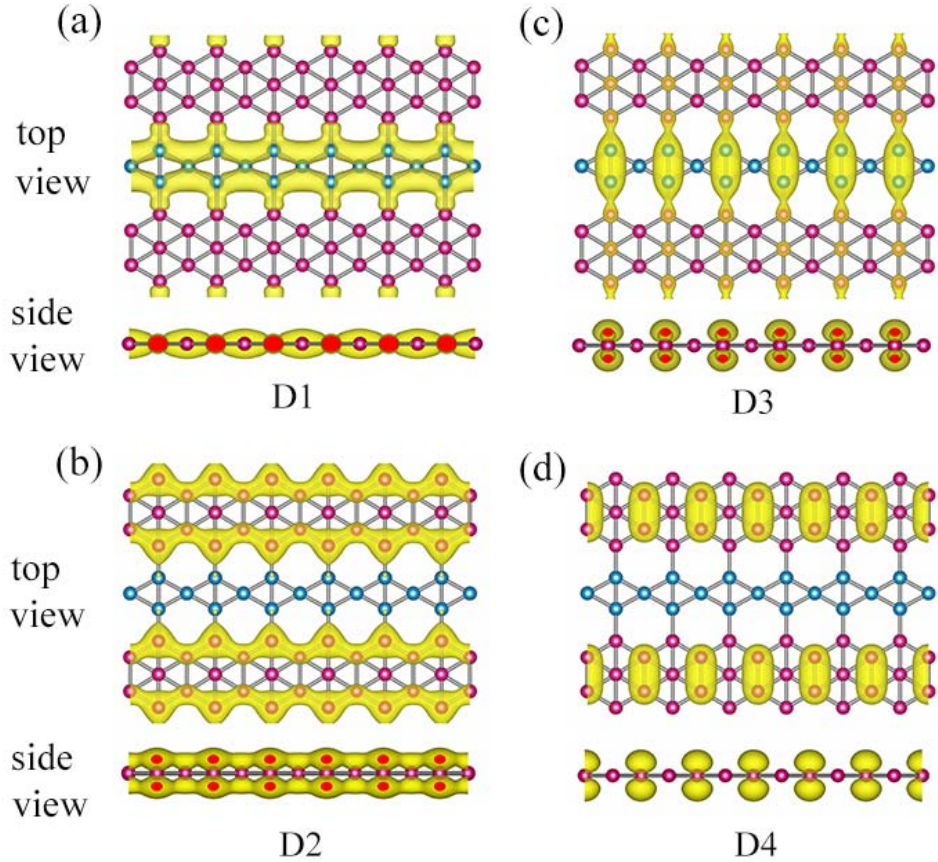

**Figure 4.** (a)-(d) Top and side views of charge densities for the states in the bands D1, D2, D3 and D4 around the Fermi level in Fig.3(a), respectively.

The crossing point along $\mathrm{X}-\mathrm{M}$ (Figure 3a) is induced by $p_z$ orbitals. The linear band D3 and flat band D4 are attributed by $p_z$ orbitals of $B_r$ and $B_h$ atoms, respectively. Checking the charge densities of states in the two bands reveals that the electrons in the D4 state are localized on junction of two adjacent hexagons, indicating that the flat band is induced by localized state. The D3 state seems also like a localized state, whose electrons are mainly localized on the B dimers in the rhombus stripes. However, these dimers can be weakly linked by the hexagon stripes, such a weak linking forms an electron channel perpendicular to the stripes and results in the linear band.

As mentioned above, the Dirac nodal lines are related to the isolated 1D channels in the rhombus and hexagon stripes. To further explain the mechanism of the 1D channels, we analyzed the chemical bonding patterns of hr-sB sheet by using the recently developed solid state Adaptive Natural Density Partitioning (SSAdNDP) method.[13] Since each B atom has 3 valence electrons, thus there are 24 valence electrons in one unit cell. According to our computations, one unit cell contains two localized two-center-two-electron (2c-2e) σ bonds, six 3c-2e σ bonds, one 4c-2e σ bond, two 6c-2e π bonds, and one 7c-2e π bond, accounting for 24 electrons per unit cell (Figures 5(a-e)). These bonds can be categorized by three kinds: in the hexagon-stripe are four 3c-2e bonds, one 4c-2e bond, and one 7c-2e bond; in the rhombus stripe are two 3c-2e bonds; two 2c-2e bonds and two 6c-2e bonds connecting the hexagon and rhombus stripes to form the overall two-dimensional structure. Note that the bonding formations in the rhombus stripe are quite different from those in the hexagon stripe. Thus, the bond lengths in the two stripes are also quite different (Figures 5(f)). All the bond lengths in the rhombus stripe are smaller than 1.67 Å, while those in the hexagon stripe are larger than 1.67 Å. These significant differences result in the isolated 1D transport channels in the two stripes.

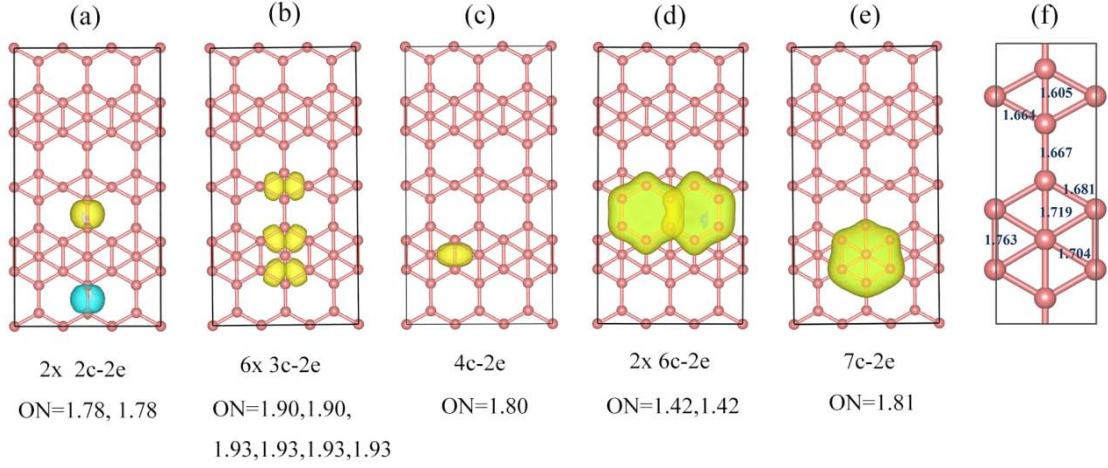

**Figure 5.** (a-e) Schematic of SSAdNDP chemical bonding patterns for hr-sB with occupation numbers (ONs). (f) Bond lengths of the monolayer sheet.

Because only the $p_z$ orbital of $B_h$ atoms and $p_{x/y/z}$ orbitals of $B_r$ atoms contribute to the states around the Fermi level, we can use a tight-binding (TB) model based on these orbitals to describe the properties around the Fermi level,

$$H = \sum_\alpha \sum_i \varepsilon_{i\alpha} + \sum_{\alpha,\beta} \sum_{i,j} t_{i\alpha,j\beta} \alpha_{i\alpha}^\dagger \alpha_{j\beta} \tag{1}$$

where $i, j \in \{1, 2, ..., 8\}$ are sites of the eight atoms in the primitive cell, $\alpha, \beta \in \{p_x, p_y, p_z\}$ are $p$ oribitals, $t_{i\alpha,j\beta}$ is the hopping energy between orbital $\alpha$ at site $i$ and orbital $\beta$ at site $j$. We only study the nearest-neighbor hopping, thus only $p_z$ orbital of $B_h$ atoms and $p_{x/y/z}$ orbitals of $B_r$ atoms are considered. The hopping parameters $t_1$ and $t_2$ are used to describe the inner interactions of rhombus stripes, $t_4$, $t_5$ and $t_6$ describe the inner interactions of hexagon stripes, while $t_3$ describes the interactions between stripes (Figure 1).

Figures 6 presents the tight-binding band structure for the hr-sB sheet, and Table 2 lists the corresponding parameters of the hopping energies. Note that the band structures obtained by tight binding and DFT computations agree well. When the parameter $t_3$ is changed from 0 to -1.3 eV, the variation has no effect on the Dirac nodal lines, but significantly affects the linear band D3 along $X - M$ (see Figure S1 in the Supporting Information (SI)). This finding further proves that the Dirac nodal lines are related to the isolated channels in the stripes, while the tilted Dirac cones are related to the interaction between stripes.

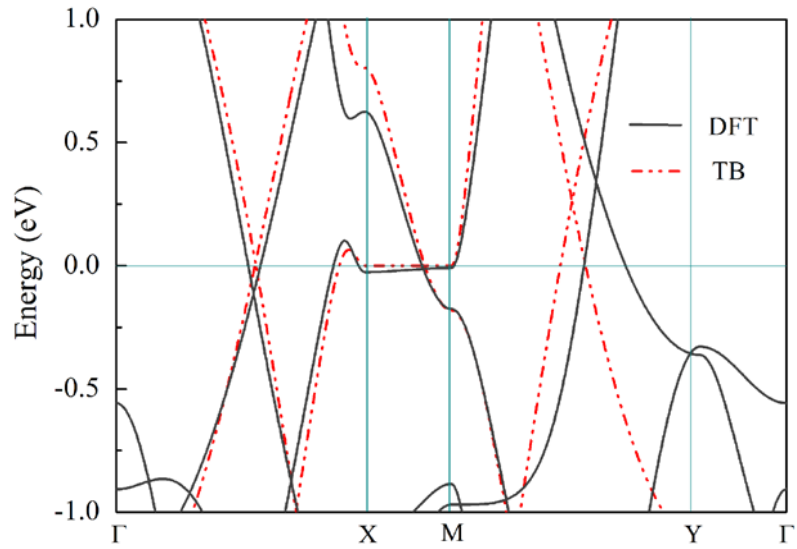

**Figure 6.** Comparison of band structures for DFT results (gray lines) and TB model in Eq. (1) (red dashed lines).

**Table 2:** The values of hopping energies $t_1 \sim t_6$ between different atomic orbitals in hr-sB. The units of all parameters are eV.

|  | $t_1$ | $t_2$ | $t_3$ | $t_4 = t_5 = t_6$ |
|---|---|---|---|---|
| $p_x p_x$ | 2.0 | 2.8 | - | - |
| $p_y p_y$ | 3.8 | 2.8 | - | - |
| $p_x p_y$ | -2.0 | -1.0 | - | - |
| $p_z p_z$ | -1.8 | -1.8 | -1.3 | -2.3 |
| On-site energy |  | $\varepsilon_{px} = \varepsilon_{py} = -3.5$ |  | $\varepsilon_{pz} = 2.3$ |

Our computations already demonstrated the high feasibility to synthesize hr-sB. However, which approach would be our first choice? Recently, Feng *et. al* successfully prepared $\beta_{12}$- and $\chi_3$- sheets by epitaxial grwoth on a Ag (111) substrate, and the hr-sB is a planar boron sheet very similar to $\beta_{12}$ and $\chi_3$. Thus, we explored the possibility to synthesize hr-sB on a Ag (111) substrate by applying an analogous experimental process. According to our computations, the lattice constants of a 1×3 supercell of hr-sB are $a' = 2.91$ Å and $b' = 25.13$ Å, and those of a 1×5√3 supercell of Ag (111) substrate are $a'' = 2.89$ Å and $b'' = 25.00$ Å. The mismatch between them are smaller than 1%. The adhesion energy between hr-sB sheet and a Ag (111) substrate (-0.131 eV/B) is comparable with those of $\beta_{12}$ (-0.141 eV/B) and $\chi_3$ (-0.116 eV/B) on a Ag (111) substrate (for more details, see Figure S2). Thus, it is rather optimistic for us to grow hr-sB on the Ag (111) substrate.

In addition, hr-sB sheet may be a good candidate for superconductor. Firstly, it has a large density of state around the Fermi level because of the flat band (Figure 3b). Secondly, there exist Dirac electrons in the sheet, corresponding to a large carrier mobility. Moreover, the Dirac points in the nodal lines distribute on the Fermi level up and down. The hole and electron pockets may be in favor of forming electron-hole pairs. Thirdly, boron is very light element. These parameters lead to a large electron-phonon coupling strength, according to $\lambda = \frac{N(0)\langle g^2 \rangle}{M\langle \omega^2 \rangle}$,[49] where $N(0)$ is the density of states at the Fermi level, $\langle g^2 \rangle$ is the average square of the electron-phonon matrix element, $M$ is the mass, and $\langle \omega^2 \rangle$ is averaged over the phonon spectrum. Therefore, the hr-sB sheet could be a good superconductor with high critical temperature.

**Conclusions**

We identified a new boron sheet, namely hr-sB, and investigated its electronic properties by first principles computations. The hr-sB monolayer consists of rhombus and hexagon strips with a hexagonal vacancy density $\eta = 1/5$. Interestingly, two types of Dirac electrons coexist in this nanosheet. One type is the electrons in the Dirac nodal lines traversing the whole BZ, whose velocities can approach $10^6$ m/s. These electrons transport along 1D channels in the stripes. The other type is the electrons in the tilted semi-Dirac cones, where strong anisotropic electronic properties are observed. The unique electronic characteristics are originated from its special bonding patterns. Some clues also indicate that hr-sB may be a good superconductor. This newly predicted boron nanosheet has a cohesive energy comparable with the experimentally available $\delta_6$-, $\beta_{12}$- and $\chi_3$- sheets, and also has outstanding dynamic and thermal stabilities, thus is highly feasible for experimental realization. To assist future experimental explorations, we proposed a promising approach to prepare hr-sB by epitaxial growth on a Ag (111) substrate.

Our studies not only found a new monolayer boron sheet, but also revealed isolated Dirac channels and new Dirac fermions which have not been observed before. We hope that these studies will inspire experimental and theoretical studies on Dirac fermions and other quasi-particles, which will help further understand the chemical bonding in boron materials, and extend the applications of boron sheet to electronics.

**Acknowledgments**

This work was supported in China by the National Science Foundation of China (Nos. 51376005 and 11474243) and in USA by Department of Defense (Grant W911NF-15-1-0650).

**Supporting Information Available:** The tight-binding band structures of hr-sB with hopping energy $t_3 = 0.0$, -0.7 and -1.3, the structure model of monolayer hr-sB sheet on Ag(111). This material is available free of charge via the internet at http://pubs.acs.org.

## REFERENCES
1. Novoselov, K. S.; Geim, A. K.; Morozov, S. V.; Jiang, D.; Katsnelson, M. I.; Grigorieva, I. V.; Dubonos, S. V.; Firsov, A. A. *Nature* **2005,** 438, 197-200.
2. Zhang, Y.; Tan, Y. W.; Stormer, H. L.; Kim, P. *Nature* **2005,** 438, 201-4.
3. Du, X.; Skachko, I.; Duerr, F.; Luican, A.; Andrei, E. Y. *Nature* **2009,** 462, 192-5.
4. Castro Neto, A. H.; Guinea, F.; Peres, N. M. R.; Novoselov, K. S.; Geim, A. K. *Rev. Mod. Phys.*

**2009,** 81, 109-162.

5. Bolotin, K. I.; Sikes, K.; Jiang, Z.; Klima, M.; Fudenberg, G.; Hone, J.; Kim, P.; Stormer, H. *Solid State Commun.* **2008,** 146, 351-355.
6. Hwang, E.; Sarma, S. D. *Phys. Rev. B* **2008,** 77, 115449.
7. Malko, D.; Neiss, C.; Görling, A. *Phys. Rev. B* **2012,** 86.
8. Huang, H.; Duan, W.; Liu, Z. *New J. Phys.* **2013,** 15, 023004.
9. Malko, D.; Neiss, C.; Vines, F.; Gorling, A. *Phys. Rev. Lett.* **2012,** 108, 086804.
10. Huang, H.; Zhou, S.; Duan, W. *Phys.Rev. B* **2016,** 94, 121117.
11. L Muechler, A. A., T Neupert. *arXiv preprint arXiv:1604.01398* **2016**.
12. M Yan, H. H., K Zhang, E Wang, W Yao. *arXiv preprint arXiv:1607.03643* **2016**.
13. Galeev, T. R.; Dunnington, B. D.; Schmidt, J. R.; Boldyrev, A. I. *Phys. Chem. Chem. Phys.* **2013,** 15, 5022-5029.
14. Chen, Y.; Xie, Y.; Yang, S. A.; Pan, H.; Zhang, F.; Cohen, M. L.; Zhang, S. *Nano Lett.* **2015,** 15, 6974-8.
15. Wu, X.; Dai, J.; Zhao, Y.; Zhuo, Z.; Yang, J.; Zeng, X. C. *ACS Nano* **2012,** 6, 7443-7453.
16. Tang, H.; Ismail-Beigi, S. *Phys. Rev. B* **2010,** 82.
17. Liu, Y.; Penev, E. S.; Yakobson, B. I. *Angew. Chem. Int. Ed. Engl.* **2013,** 52, 3156-9.
18. Penev, E. S.; Bhowmick, S.; Sadrzadeh, A.; Yakobson, B. I. *Nano Lett.* **2012,** 12, 2441-5.
19. Zhang, Z.; Penev, E. S.; Yakobson, B. I. *Nat. Chem.* **2016,** 8, 525-7.
20. Özdoğan, C.; Mukhopadhyay, S.; Hayami, W.; Güvenç, Z. B.; Pandey, R.; Boustani, I. *J. Phys. Chem. C* **2010,** 114, 4362-4375.
21. Zhang, Z.; Xie, Y.; Peng, Q.; Chen, Y. *Nanotech.* **2016,** 27, 445703.
22. Tang, H.; Ismail-Beigi, S. *Phys. Rev. Lett.* **2007,** 99, 115501.
23. Liu, H.; Gao, J.; Zhao, J. *Sci. Rep.* **2013,** 3, 3238.
24. Penev, E. S.; Kutana, A.; Yakobson, B. I. *Nano Lett.* **2016,** 16, 2522-6.
25. Mannix, A. J.; Zhou, X.-F.; Kiraly, B.; Wood, J. D.; Alducin, D.; Myers, B. D.; Liu, X.; Fisher, B. L.; Santiago, U.; Guest, J. R.; Yacaman, M. J.; Ponce, A.; Oganov, A. R.; Hersam, M. C.; Guisinger, N. P. *Science* **2015,** 350, 1513-1516.
26. Feng, B.; Zhang, J.; Zhong, Q.; Li, W.; Li, S.; Li, H.; Cheng, P.; Meng, S.; Chen, L.; Wu, K. *Nat. Chem.* **2016,** 8, 563-8.
27. Xu, L. C.; Wang, R. Z.; Miao, M. S.; Wei, X. L.; Chen, Y. P.; Yan, H.; Lau, W. M.; Liu, L. M.; Ma, Y. M. *Nanoscale* **2014,** 6, 1113-8.
28. Su, C.; Jiang, H.; Feng, J. *Phys. Rev. B* **2013,** 87.
29. Wang, Z.; Zhou, X. F.; Zhang, X.; Zhu, Q.; Dong, H.; Zhao, M.; Oganov, A. R. *Nano Lett.* **2015,** 15, 6182-6.
30. Katsnelson, M. I. *Mater. today* **2007,** 10, 20-27.
31. Brunetto, G.; Autreto, P. A. S.; Machado, L. D.; Santos, B. I.; dos Santos, R. P. B.; Galvão, D. S. *J. Phys. Chem. C* **2012,** 116, 12810-12813.
32. Yin, W.-J.; Xie, Y.-E.; Liu, L.-M.; Wang, R.-Z.; Wei, X.-L.; Lau, L.; Zhong, J.-X.; Chen, Y.-P. *J. Mater. Chem. A* **2013,** 1, 5341.
33. Liu, Y.; Wang, G.; Huang, Q.; Guo, L.; Chen, X. *Phys. Rev. Lett.* **2012,** 108, 225505.
34. Wang, J.; Huang, H.; Duan, W.; Liu, Z. *J. Chem. Phys.* **2013,** 139, 184701.
35. Zhong, C.; Chen, Y.; Xie, Y.; Yang, S. A.; Cohen, M. L.; Zhang, S. B. *Nanoscale* **2016,** 8, 7232-9.

36. Piazza, Z. A.; Hu, H. S.; Li, W. L.; Zhao, Y. F.; Li, J.; Wang, L. S. *Nat. Commun.* **2014,** 5, 3113.
37. Zhou, X.-F.; Dong, X.; Oganov, A. R.; Zhu, Q.; Tian, Y.; Wang, H.-T. *Phys. Rev. Lett.* **2014,** 112.
38. Ma, F.; Jiao, Y.; Gao, G.; Gu, Y.; Bilic, A.; Chen, Z.; Du, A. *Nano Lett.* **2016,** 16, 3022-8.
39. Jiao, Y.; Ma, F.; Bell, J.; Bilic, A.; Du, A. *Angew. Chem. Int. Ed. Engl.* **2016,** 55, 10292-5.
40. Lopez-Bezanilla, A.; Littlewood, P. B. *Phys. Rev. B* **2016,** 93.
41. Zhang, R.; Yang, J. *arXiv preprint arXiv:1607.08290* **2016**.
42. Xu, L.; Du, A.; Kou, L. *arXiv preprint arXiv:1602.03620* **2016**.
43. Kresse, G.; Hafner, J. *Phys. Rev. B* **1993,** 47, 558.
44. Perdew, J. P.; Chevary, J.; Vosko, S.; Jackson, K. A.; Pederson, M. R.; Singh, D.; Fiolhais, C. *Phys. Rev. B* **1992,** 46, 6671.
45. Perdew, J. P.; Burke, K.; Ernzerhof, M. *Phys. Rev. Lett.* **1996,** 77, 3865-3868.
46. Kresse, G.; Joubert, D. *Phys. Rev. B* **1999,** 59, 1758.
47. Monkhorst, H. J.; Pack, J. D. *Phys. Rev. B* **1976,** 13, 5188-5192.
48. Togo, A.; Oba, F.; Tanaka, I. *Phys. Rev. B* **2008,** 78.
49. McMillan, W. L. *Phys. Rev.* **1968,** 167, 331-344.